\documentclass[10pt,conference]{IEEEtran}

\usepackage{layout}
\usepackage{graphicx}
\usepackage{amsmath}
\usepackage{amssymb}
\usepackage{amscd}
\usepackage{times}

\usepackage{multirow}
\newcommand{\mr}[1]{\multirow{2}*{#1}}
\newcommand{\fourr}[1]{\multirow{4}*{#1}}
\newcommand{\threer}[1]{\multirow{3}*{#1}}

\begin{document}

\title{{\it IMPACT}: {\it I}nvestigation of {\it M}obile-user {\it P}atterns {\it A}cross University {\it C}ampuses using WLAN {\it T}race Analysis}
\author{\authorblockN{Wei-jen Hsu and Ahmed Helmy}
\authorblockA{\\Department of Electrical Engineering, University of Southern California
\\Email: \{weijenhs, helmy\}@usc.edu}}

\maketitle
\bibliographystyle{IEEEtran}
\begin{abstract}
We conduct the most comprehensive study of WLAN traces to date. Measurements collected
from four major university campuses are analyzed with the aim of developing
fundamental understanding of realistic user behavior in wireless networks. Both
individual user and inter-node (group) behaviors are
investigated and two classes of metrics are devised to capture the underlying
structure of such behaviors. 

For individual user behavior we observe distinct patterns in which most users are 'on'
for a small fraction of the time, the number of access points visited is very small
and the overall on-line user mobility is quite low. We clearly identify categories of
heavy and light users. In general, users exhibit
high degree of similarity over days and weeks. 

For group behavior, we define metrics for encounter patterns and friendship.
Surprisingly, we find that a user, on average, encounters less than 6\% of the
network user population within a month, and that encounter and friendship relations
are highly asymmetric. We establish that number of encounters follows a biPareto
distribution, while friendship indexes follow an exponential distribution. We capture
the encounter graph using a small world model, the
characteristics of which reach steady state after only one day.

We hope for our study to have a great impact on realistic modeling of network
usage and mobility patterns in wireless networks.
\end{abstract}

\section{Introduction}

Recently, wireless networks have been gaining popularity and are being deployed ubiquitously in various environments, especially on university campuses. With more users switching to wireless networks, the importance of understanding user behavior in such environments is increasing. First, analysis of user behavior and network usage patterns enables accurate assessment of wireless network utilization and aids in developing better management techniques and capacity planning decisions. Second, usage analysis is also a necessary first step towards developing realistic models of usage patterns and mobility models that are crucial for the design and evaluation of wireless networking protocols. Third, as new technologies evolve (e.g. variants of $802.11$ WLANs, or ad hoc networks), fundamental understanding of user behavior becomes essential for the successful deployment of the emerging technology. 

Although many wireless network protocols have been developed over the past decade, the majority have been designed independently of the context in which they may be deployed and are usually evaluated using artificial (e.g.
synthetic, often unrealistic) models. We believe that the design and evaluation of the next generation wireless networks should go hand-in-hand with deep, insightful understanding of the realistic environments in which they will be deployed and used. 

The main focus of this paper is to gain further understanding of realistic user \footnote{In this paper we use the terms "user", "node", and "mobile node (MN)" interchangeably to describe a wireless network user.} behavior (e.g. usage, mobility) utilizing the most extensive wireless LAN traces collected to date from four major university campuses. Our study is different from (but relates to) studies on {\it mobility modeling} \cite{RWP-harmful}, \cite{stationarity-model} . The WLAN traces provide {\it coarse-grained} location snapshots of users. In this work we seek to describe user behavior at access point (AP) or building granularity. While mobility models describe how users move, the WLAN trace captures the combined result of movement, network access and usage patterns of users. In that sense, one may envision that all-encompassing models may be built using our study (but perhaps not pure mobility models). Such models may be used to validate mobility models at a coarse-grained level. While it is difficult to collect large amount of raw mobility traces, existing university WLANs provide a very good opportunity to collect WLAN usage traces and enable analysis of user behavior.

A few studies have been previously conducted on WLAN traces \cite{MIT-trace}, \cite{UCSD-trace}, \cite{Dart-trace}, and we do borrow from these traces and studies as appropriate. These studies are quite helpful, but each of them was conducted on a single campus, and hence it becomes unclear whether their findings generalize beyond the studied campus. Furthermore, most of these studies focus on metrics for individual users and building or access point (AP) usage. In our study, we go beyond previous work to compare user behavior across different campuses, which allows us to generalize our findings and reason about commonalities and differences between campuses. Also we define new metrics and models to study group behavior, in addition to individual users, to capture very important inter-user correlations. Our approach enables us to obtain new (sometimes surprising) results on the fundamental understanding of wireless network usage. Hence, this study is quite novel in various aspects and we expect it to greatly impact future research on realistic and trace-based modeling and analysis of wireless networks.

In this paper we propose two categories of metrics to analyze the behavior of individual user and groups of users. From the individual user analysis, we observe that exact user behaviors are different not only due to the underlying environment and campus, but also because of various methods used for trace collection and analysis. In general, one can identify categories of heavy users and light users, and each user tends to overall use a very small subset of the APs on campus. In most cases we identify repetitive patterns in user behavior over various time frames (e.g., days, weeks). We further study the {\it inter-node relationship between users} to understand group behavior. By looking at {\it encounters} relationships among users and defining {\it friendship indexes} we provide various new methodologies to understand underlying user behaviors in wireless networks. The distribution of both encounters and friendship indexes are highly asymmetric, indicating a heterogeneous user population. 
Surprisingly, we find that a user, on average, encounters between 1.8\% and 6\% of the network user population within a month. We establish that number of encounters follows a biPareto distribution, while friendship indexes follow an exponential distribution. We utilize a Small World model to understand the relationship in the encounter graphs formed by wireless network users. To our surprise the metrics of the formed Small Worlds saturate after only one day (in a one-month trace period). Finally, we propose information diffusion experiments to reveal richness of encounter patterns. Many of our findings point to invalid assumptions often made in mobility modeling and simulation and provide guidelines for realistic modeling of user behavior on university campuses.

The major contributions of this paper are:

\begin{table*}
\caption{Statistics of studied traces}
\scriptsize
\label{trace-facts}
\begin{center}
\begin{tabular}{|c||c|c|c|c|c|c|c|c|c|c|}
\hline
Trace & Unique & Unique & Unique & Trace & \mr{User type} & \mr{Environment} & Trace collection & Analyzed part & Users in & Labels used \\
 source &  users & APs & buildings & duration & & & method & in this paper & analyzed part & in graphs \\
\hline
\mr{MIT\cite{MIT-trace}} & \mr{1,366} & \mr{173} & \mr{3} & Jul. 20 '02 to & \mr{Generic} & 3 Engineer & \mr{Polling} & \mr{Whole trace} & \mr{1,366} & MIT-cons \\
& & & & Aug. 17 '02 & & buildings & & & & MIT-rel \\
\hline
\fourr{Dartmouth\cite{Dart-trace}} & \fourr{10,296} &\fourr{623} & \fourr{188} & & \fourr{Generic} & & \fourr{Event-based} & Jul. 2003 & 2,518 & Dart-03 \\
\cline{9-11}
& & & & Apr. '01 to & & Whole & &\threer{Mar. 2004} & \threer{5,416} & Dart-04\\
& & & & Jun. '04 & & campus & & & & Dart-rel \\
 & & & & & & & & & &  Dart-cons \\
\hline
\mr{UCSD\cite{UCSD-trace}} & \mr{275} & \mr{518} & \mr{N/A} & Sep. 22 '02 to & \mr{PDA only} & Whole & \mr{Polling} & Sep. 22 '02 to & \mr{275} & \mr{UCSD} \\
& & & & Dec. 8 '02 & & campus & & Oct. 21 '02 & &\\
\hline
\mr{USC} & \mr{4,548} & 79 & \mr{73} & Dec 03-Now (trap) & \mr{Generic} & Whole & \mr{Event-based} & Apr.
20, '05 to & \mr{4,528} & \mr{USC}\\
& & ports & & Apr 20 05-Now (detail) & & campus & & May. 19 '05 & &\\
\hline
\end{tabular}

\end{center}
\end{table*}

\begin{itemize}
\item By using WLAN traces from four different campuses, comparing the results and highlighting both similarities and differences, it is the largest scale trace-based study in the literature as far as we know.

\item By proposing metrics for describing individual MN behaviors, we construct a basis on which models for individual MNs can be established. We also find several facts that indicate traditional, randomly generated synthetic mobility models (such as random waypoint, random walk, etc.) are not adequate for a heterogeneous environment such as university campuses.

\item By proposing new methodologies to understand relationships between MNs, we build tools and understandings that should be useful for future research.

\end{itemize}

The rest of the paper is organized as follows: In section \ref{RW} we discuss the related works. The studied wireless network environments and trace-collection related issues are discussed in section \ref{environment}. We introduce various metrics to describe individual user behavior in-depth in section \ref{user-metric}. After that, we study the relationship between MNs by designing various experiments in section \ref{user-relationship}. Finally, we provide some discussions and directions for potential future work in section \ref{disc} and conclude the paper in section \ref{conclusion}.

\section{Related work} \label{RW}

Influenced by the gaining popularity of wireless LANs in recent years, there are increasing interests on studying usage of wireless LANs. Several previous works \cite{MIT-trace}, \cite{UCSD-trace}, \cite{Dart-trace} have provided extensive study on wireless network usage statistics and made their traces available to the research community. Our work is built upon these understandings and traces.

With these traces available, more recent research works focus on modeling user behaviors in wireless LANs. In \cite{traffic-model} the authors propose models to describe traffic flows generated by wireless LAN users, which is a different focus to this paper. In the first part of this paper we focus more on identifying metrics that capture important characteristics of user association behaviors. We understand user associations as coarse-grained mobility at per access point granularity. Similar methodology has been used in \cite{MIT-trace} and \cite{WLAN-model}. In \cite{WLAN-model} the authors propose a mobility model based on association session length distribution and AP preferences. However, there are also other important metrics that are not included, such as user on-off behavior and repetitive patterns. We add these metrics to provide a more complete description for user behaviors in wireless networks.

Through this work, we also establish that although the general behavior of users are similar in previously collected traces, the detailed distributions vary due to differences in underlying user population and/or trace collection methods. Hence, conclusion drawn from one trace sometimes may not be generalized to describe all network environments.

Recent research works on protocol design in wireless networks usually utilize synthetic, random mobility models for performance evaluation \cite{mobility-model}, such as random waypoint model or random walk model. MNs in such synthetic models are always on and homogeneous in their behavior. Both of these characteristics are not observed in real wireless traces. We argue that to better serve the purpose of testing new protocols, we need models that capture on-off and heterogeneous behavior we observed from the traces.

There are few work on understanding relationship between users in wireless LANs in current research literature. However, such understanding is crucial for further research on the next generation (socially-aware or context-aware) protocols. In the second half of this paper we take a first step toward this end by proposing new methodologies and experiments to understand relationship between MNs in wireless LANs.

\section{Target environment and trace collection methods} \label{environment}

In this study we mainly focus on wireless traces collected from university campuses. We obtain wireless traces from
four different universities, including totally over 12,000 distinct users and over 1,300 APs. To our best knowledge this is the most extensive study of user behavior in wireless networks so far.

The traces have been collected in the studied campuses using different trace-collection methods. We summarize the important characteristics of these traces in Table \ref{trace-facts} and explain the major issues below.

The focus of the paper is on understanding the behavior of mobile nodes (MNs), including association to APs, mobility,
repetitive pattern, encounter, and friendship. These four traces are chosen to represent different campus environments, user populations, location granularity, and trace-collection methods. We study the differences and similarities of user behavior in these traces, and try to attribute them to the underlying differences in the traces as appropriate. In order to make the results we get below comparable between traces, we only analyze selected one-month chunks from the longer Dartmouth and UCSD traces. All these traces, except UCSD trace, collect measurements of generic wireless network users with various devices, including but not limited to laptops, PDAs, and VoIP devices \cite{Dart-trace}. UCSD trace is from a specific study about PDA users. All the traces, except MIT trace, are collected from the entire campus wireless network. MIT trace is collected from three engineering buildings, hence its user population is not as diverse as the other traces, and the geographic scope of trace collection is smaller. USC trace is the only one that has coarser, per switch port location granularity (approximately correspond to buildings on campus), while the others have per AP location granularity. These distinctions in underlying environments may lead to differences in the metrics discussed below. 

The methods of collecting wireless network traces can be categorized into two major categories: (i) Polling-based
methods which record the association of MNs at periodic time intervals, using SNMP \cite{MIT-trace} or association
tracking software on the MNs \cite{UCSD-trace}, and (ii) Event-based methods which record MN online/offline events using logging server (e.g. syslog) \cite{Dart-trace}. It is generally accepted that event-based approach provides more accurate records of MN behavior in the network. However, there is no in-depth study to quantify the differences between these two approaches. In order to further understand the effects of using different methods of trace collection on the trace obtained, we also obtain a {\it re-constructed polling trace} as follows: For an event-based trace, we observe the trace at regular time intervals and emulate what would be recorded if the trace is taken by polling-based method. We then process the re-constructed polling trace as we do to a normal polling-based trace, and compare the findings with the corresponding findings from the original event-based trace. We use March 2004 Dartmouth trace ({\it Dart-04}) to carry out this experiment, obtaining {\it Dart-cons} and {\it Dart-rel} traces based on the conservative and relaxed assumptions detailed below.  

For traces using polling-based approach, duration of association must be derived from the observations made at constant intervals based on an important assumption of association duration for each observed data point. We test two different assumptions in this aspect: (a) A conservative ({\it MIT-cons, Dart-cons}) approach, in which a MN is assumed with the AP only until the next expected polling (recording) epoch, unless indicted otherwise by new samples in the trace. This approach reflects what is observed from the trace faithfully, but may have the drawback that inaccuracy in polling intervals or lost SNMP records will lead to the conclusion that the MN is switching between online and offline status while it has been always on. (b) A more relaxed approach ({\it MIT-rel, Dart-rel}), in which a MN is assumed with the AP for four polling intervals after it is observed with the AP, unless indicted otherwise by the trace. This approach is more robust to disturbances in trace collection, however, it may erroneously increase the duration of association with APs after a MN is in fact offline. We use only the relaxed approach to UCSD trace.

\section{Analysis of individual user behavior} \label{user-metric}

In this section we use metrics to describe and compare behaviors of individual users (or MNs) in the studied environments. These metrics can be divided into four major categories as follows: (a) {\it Activeness of users}: This category captures the frequency of user participation in network activity. In general wireless network users are not always on, but show up in the trace intermittently. (b) {\it The long-term mobility of users}: This category captures how widely a MN moves in the network in the long run (i.e., for the whole duration of the trace), and how MNs online time is distributed among the APs. The intention here is to capture the tendency for a MN to visit various locations in the studied environment.  (c) {\it The short-term mobility of users}: This category captures how MNs move in the network while it keeps associated with some AP. The intention here is to capture the mobility of a MN while {\it using} the wireless network. (d) {\it The repetitive association pattern of users}: This category captures the user {\it on-off} behavior with respect to time of the day and the location. We expect users tend to show repetitive structure in their association patterns and propose {\it network similarity index (NSI)}, a quantitative metric to capture such repetitive pattern in user behavior.

Before presenting the analyses using the above metrics we first introduce some terminologies: 

\begin{itemize}

\item \textbf{Online event} is defined as the event when a MN associates itself with an AP, while it is not associated
with any AP right before the online event. 

\item \textbf{Offline event} is defined as the event when a MN disassociates itself with the current AP to which it is
associated, and does not associate itself with any other AP right after the disassociation event. 

\item \textbf{Handoff event} is defined as a MN changes its association from one AP to another with no time gap in
between (i.e., by issuing a re-associate at the second AP). 

\item \textbf{Association session} is defined as the duration between an online event to the next offline event for the
same MN. Handoff events do not terminate an association session.

\item \textbf{Total online time} is the sum of time periods a MN associated with any AP throughout the studied trace. 

\item \textbf{Existence time} is the time difference between a MN's first online event and its last offline event in
the studied trace. It is a conservative measure of the time duration for which the MN is a
potential user of the network.

\end{itemize}

\subsection{The activeness of the users}

Activeness of users is the first aspect we look into in attempt to compare the different traces. Activeness of users can be captured by either total online time fraction of a MN or the number of association sessions generated by a MN. 

\begin{figure}
\centering
\includegraphics[width=2.6in]{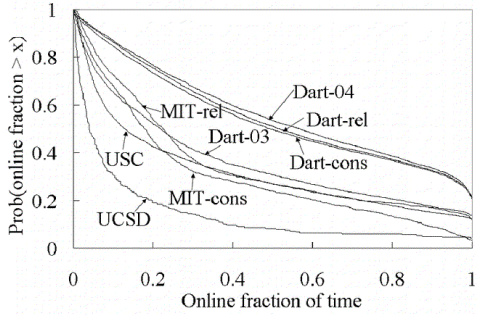}
\caption{CCDF of online time fraction}
\label{CCDF-ontime}
\end{figure}

Online time fraction is not straight forward to measure due to users joining or leaving the network. We choose to define the {\it online time fraction} as the ratio between MN's total online time to its existence time. We plot the CCDF of online time fraction of users in various traces in Fig. \ref{CCDF-ontime}. 

From Fig. \ref{CCDF-ontime} we observe that in all traces only a small protion of users are always on. Even for the
most active Dartmouth trace (Dart-04), there are only less then $30\%$ always-on users. These observations argue
strongly that most users have {\it on-off} usage patterns, where some of the users are {\it heavy} users (with high {\it on} time) while many are {\it light} users. The distributions of the {\it on/off} times seem to depend heavily on the environment (i.e., campus). UCSD trace, which focused only on PDA users, is the least active one among all traces. The other traces (MIT, USC, Dart-03) are not very different in online time fraction distribution. The activeness of MNs increased from 2003 to 2004 in Dartmouth trace, which agrees with the findings in \cite{Dart-trace}. By comparing the curves of Dart-04, Dart-rel, and Dart-cons, we observe that online time fraction is consistent for the same trace under different trace collection (or trace reconstruction) methods. Hence, this metric is insensitive to the tracing method.

\begin{figure}
\centering
\includegraphics[width=2.6in]{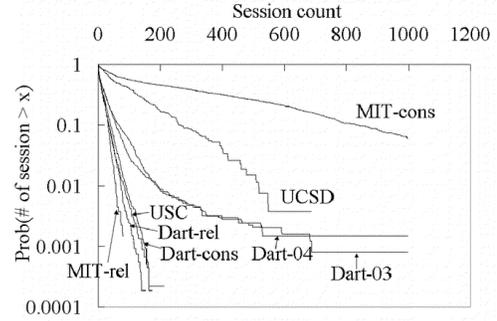}
\caption{CCDF of number of association sessions by users}
\label{CCDF-session}
\end{figure}

We further compare the CCDF of number of association sessions generated by users in these traces in Fig. \ref{CCDF-session}. We observe that the PDA users in UCSD trace generate more association sessions than users in other traces, which are generic wireless network devices (mainly laptop users) during comparable trace duration. This fact, together with the less online time fraction in Fig. \ref{CCDF-ontime}, indicates that PDA users are more likely to use the devices for shorter but more frequent sessions. From the figure we also observe that count of association sessions is sensitive to the trace collection method. While the two original Dartmouth traces (Dart-04 and Dart-03) show similar distributions, the re-constructed traces (Dart-rel and Dart-cons) show very different distributions form the original Dart-04 trace, since traces collected by polling at regular intervals will overlook association sessions shorter than the polling interval. Another technical difficulty here is to adequately translate a record seen in polling-based traces to the duration of association. As we compare the curves of MIT-cons and MIT-rel, we find them drastically different. A closer investigation reveals that in the MIT trace although SNMP polling intervals are typically 5 minutes apart, sometimes records of MN association are obtained at longer intervals, and this leads to bogus terminations and re-initiation of association sessions if the conservative assumption is used, leading to the high association session counts shown by curve MIT-cons. 

{\it \textbf{On-off behavior is very common for wireless users. This seems especially true for small
handheld devices. There are clear categories of heavy and light users, the distribution of which is skewed and
heavily depends on the campus. The 'number of sessions' metric is sensitive to the trace collection method, while
the 'online time fraction' is insensitive to such method.}}

\subsection{The long-term mobility of users}

In this section we capture the long-term mobility of users by obtaining the overall statistics of AP association history during the whole trace period. We investigate the number of APs a user associates with and the fraction of online time it associates with each of the APs. 

\begin{figure}
\centering
\includegraphics[width=2.6in]{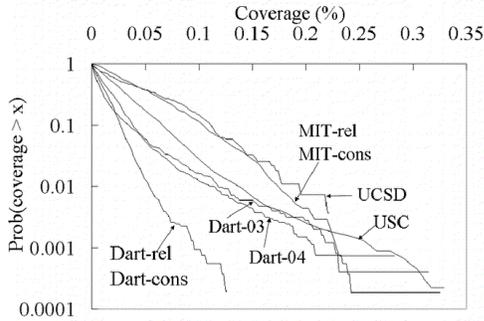}
\caption{CCDF of coverage of users}
\label{CCDF-coverage}
\end{figure}

We define {\it coverage} of a user as percentage of APs it associates with during the trace period, which is the
number of APs the node have associated with over the total AP number in the trace.  For USC trace we use switch ports (approximately corresponding to buildings) in place of APs. The distributions of coverage of users in the traces are shown in Fig. \ref{CCDF-coverage}. This metric captures how widely a user moves for the whole period of trace in the studied network environment.

We observe that users have small coverage in all environments. None of them have a user visiting more than $40\%$ of all APs. The MIT trace focuses on only three buildings, hence the relative coverage of users is much higher. In UCSD trace, the PDA users seem likely to visit a larger portion of campus than the generic users do in the other campus-wide traces, seemingly due to the portability of PDAs. Coverage seems to remain stable with respect to time change (compare Dart-03 and Dart-04), but it is sensitive to the trace collection method since the polling-based method overlooks short sessions and under-estimates the coverage metric. However, different re-construction methods of the polling-based trace (con, rel) result in the same coverage, as the metric counts the number of APs a MN associates with, not the association duration. 

\begin{figure}
\centering
\includegraphics[width=2.6in]{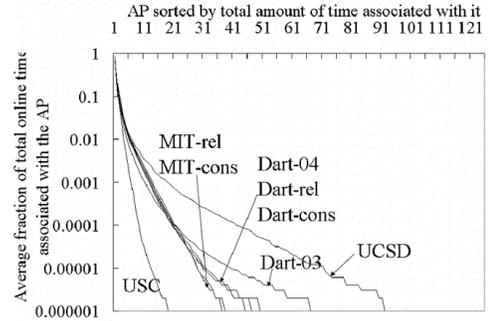}
\caption{Average fraction of time a MN associated with APs. For each MN, the AP list is sorted based on prevalence
values before taking average}
\label{time-frac-AP}
\end{figure}

In \cite{MIT-trace} the authors define {\it prevalence} as the fraction of online time a MN spends associated with each of the APs during the trace period. We follow that definition and compare the distribution of prevalence across different traces as follows: In order to understand how a user distribute its total online time among the APs it has association with, we order the APs by the prevalence value for each the MN, and take average of prevalence values across all MNs for the same AP ranking to get the curves showing average association time fraction in Fig. \ref{time-frac-AP}. 

From Fig. \ref{time-frac-AP} we observe that for all environments, the general trend is that each user has very few APs at which it spends most of its online time. In particular, for all the traces on average a MN spends more than $65\%$ of its online time with {\it one} AP, and more than $95\%$ of online time at as few as $5$ APs. The left-end of the curves are similar, but the tail varies. The higher mobility of PDA users in UCSD trace translates into a longer tail, where in addition to those few most frequent APs, the users also access the wireless network at much more locations with small time fraction as compared to other traces. This metric is robust to different trace collection methods and assumptions of trace translation, as the curve for Dart-04 is close to Dart-cons or Dart-rel. Same for MIT traces. 

{\it \textbf{Individual users access only a very small portion of APs in the network, less than $40\%$ in all campuses. The long-term mobility of users displays strong skewness of time associated with each AP. On average a user spends more than $95\%$ of time at its top five most visited APs.}}

\subsection{The short-term mobility of users}

\begin{figure}
\centering
\includegraphics[width=2.6in]{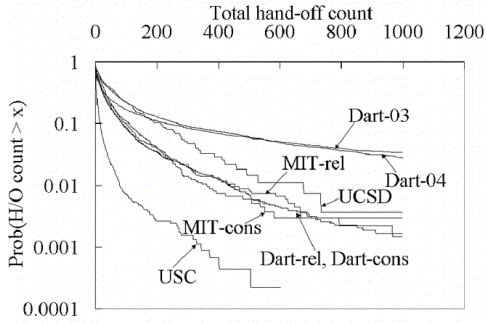}
\caption{CCDF of total handoff count per MN}
\label{CCDF-HO}
\end{figure}

\begin{figure}
\centering
\includegraphics[width=2.6in]{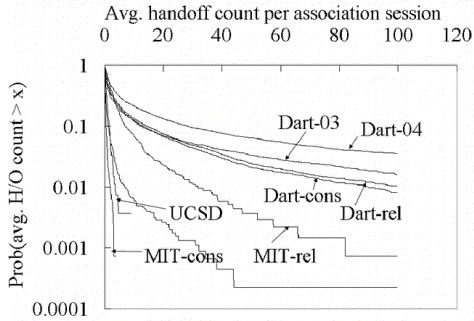}
\caption{CCDF of average handoffs per association session}
\label{CCDF-HO-session}
\end{figure}

In this section we study the per-association session mobility of a user, which reflects their short-term mobility. This captures a different dimension of a user as the previous section: How mobile the user is while {\it using} the network? We use handoff statistics as a measure of user mobility while using the network, looking at both distributions of the total number of handoffs and the average number of handoffs per association session of each user. We plot the curves in Fig. \ref{CCDF-HO} and \ref{CCDF-HO-session}, respectively.

Our first intuition is that user mobility should be dependent on the device type, and PDAs in UCSD trace should display higher mobility than users in other traces. However, as shown in both figures, the UCSD trace does not show more handoff counts than other traces. On the contrary, it is among the least in average handoff per association session. This may be related to the fact that PDAs are usually used for short durations, hence experience less handoff events. The exact number of handoff count depends heavily on the network environment. In USC trace, the coarse location granularity directly leads to smaller handoff counts. On the other hand, Dartmouth traces have much more average handoff counts per session than the other traces. 

We also observe that handoff counts in Fig. \ref{CCDF-HO} are sensitive to the trace collection method, as curve for Dart-04 differs significantly from Dart-rel and Dart-cons. This is again because the polling-based method overlooks short association sessions and hence many hand-off events are not captured. 

From Fig. \ref{CCDF-HO-session} we also observe that less than $20\%$ of users have more than $10$ handoffs per association session in all traces. This implies that most users are relatively stationary while using wireless devices. Also, the handoff statistics presented here are subjected to ping-pong effect \cite{Dart-trace}, referring to excessive handoff events due to disturbance in wireless channels while the MN itself might be stationary. Hence, we expect the actual short-term mobility of users is even lower than the results we get from the traces directly.

{\it \textbf{The majority of users experience low mobility while using the network. This is even true for portable devices such as PDAs. The actual handoff statistics depend heavily on the environment.}}

\subsection{The repetitive association pattern of users}

Naturally user behavior changes with respect to time of the day and day of the week, as people follow daily and weekly schedules in their lives. In some cases, user association pattern repeats itself day to day or week to week. In this section we try to quantify such repetitive pattern by defining the {\it network similarity index (NSI)} below.

We start the definition with {\it location similarity index} for individual users. First we take snapshots of associated APs of the user every 1 minute. To study the tendency of the user showing repetitive behavior after a certain time gap (for example, every 24 hours), we consider all snapshot pairs that are separated by this time gap, and calculate "the fraction of all such pairs where the user is associated with the same AP in both snapshots". This is an indication of how likely this user re-appears at the same location after the chosen time gap. We consider only those snapshots that fall within the existence time of the user. {\it Network similarity index (NSI)} at a given time gap is the average of {\it location similarity index} for all users at this time gap.

\begin{figure}
\centering
\includegraphics[width=2.6in]{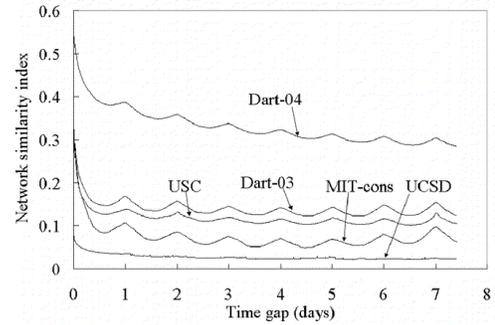}
\caption{Network similarity indexes. The peaks represent intervals for which there is high similarity.}
\label{NSI}
\end{figure}   
 
In Fig. \ref{NSI} we show the {\it NSI} for all the traces. In most of these traces we observe obvious higher network similarity index if the time gap is integer multiples of a day. This is an indication that users have the strongest tendency to show repetitive association pattern at the same time of each day. It is also interesting to observe that for USC, MIT, and Dart-03 traces, the network similarity index for the gap of 7 days (i.e., a week) is the second highest, only slightly lower than that of for the gap of 1 day. This indicates weekly repetitive pattern is also strong in these traces. On the other hand, UCSD trace shows little repetitive pattern as there is almost no obvious spikes in its {\it NSI} curve. This can be attributed to its user population being PDA users. Unlike laptops, which are more related to work, PDAs are usually used in a more casual way in short, scattered durations. Hence it is expected that PDA users show less repetitiveness in their usage pattern.

The "average value" of the {\it NSI} curves reflects the fraction of users that always stay at the same location. In that sense, we see that Dartmouth trace has the most stationary users. This may be attributed to the fact that Dartmouth traces include users in student dormitories, which are mainly stationary users and have high location similarity indexes. USC has not deployed WLAN in dormitories yet, and MIT trace is mainly focused on buildings for work.

For Dart-04 traces we observe significantly higher values in {\it NSI} curve. Another interesting distinction between Dart-04 trace and the others is that Dart-04 trace does not show a second peak in network similarity index at 7-days gap. Instead, network similarity index decreases as time gap increases. A closer investigation into Dartmouth University calendar reveals that they switch from winter quarter to spring quarter in the middle of March. We suspect that the decreasing NSI with respect to time gap might be attributed to people deviating from normal daily/weekly schedules at the end or beginning of a quarter.

{\it \textbf{We observe clear repetitive patterns of association in wireless network users. Typically, user association patterns show the strongest repetitive pattern at time gap of one day and the second strongest at one week.}}

\section{Relationship between nodes} \label{user-relationship}

In addition to individual user behavior studied in the previous section, observing relationship between MNs is also important to understand the characteristics of the traces. In this section we first investigate the distributions of {\it encounters} between MNs. Then we propose metrics to capture {\it closeness} (e.g. friendship) among MNs. Considering encounters as a way to build up {\it relationship} between MNs, we study how MNs form a relationship network via encounters by defining {\it encounter-relationship graph} (ER graph) to observe clustering and degree of separation among MNs and contrast it with the SmallWorld model \cite{Watts-SmallWorld}. Finally, we carry out simulations to test the potential of information dissemination based on encounters alone. All these experiments serve as vehicles to help us further understand the underlying structures of these traces.

\subsection{Encounters between Nodes} \label{encounter}

Nodal encounters in mobile network are important events as they provide opportunities for involved nodes to build up some relationship or to communicate directly. Here we define an {\it encounter event} as the duration of two MNs associate with the same AP during overlapping time intervals. The wireless LAN traces provide sequences of AP (switch port for USC trace) association history for MNs in the network. We can derive when MNs encounter with each other by simply comparing individual association traces. The distribution of these encounter events is the first step to understand the structure of MN relationship in the traces. The direct questions to ask about the encounter events are: What is the proportion of other nodes a typical node meet? Do nodes meet with each other repeatedly or not?

\begin{figure}
\centering
\includegraphics[width=2.6in]{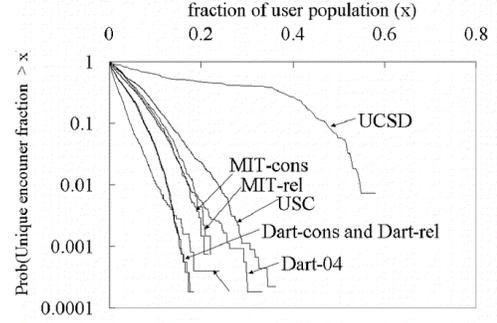}
\caption{CCDF of unique encounter fraction}
\label{UE}
\end{figure}

Fig. \ref{UE} shows the CCDF of fraction of MNs a given MN has encountered through the whole trace period. From the figure we observe that all the nodes encounter only less than $40\%$ of the user population within a month, with UCSD trace being the only exception. This may be partly due to the fact that the $275$ PDA users in UCSD trace were all selected from freshman class, and they tend to stay in several common dorms as stated in \cite{UCSD-trace}. In all the other traces, on average a MN encounters with only $1.88\%$ (Dart-03) to $5.94\%$ (USC) of the whole user population within the 30-day trace period. The rather limited population of encounter stems naturally from the fact that a MN will not visit a large portion in the network, as shown in previous section in Fig. \ref{CCDF-coverage}, and different MNs have different locations to visit on regular basis.

\begin{figure}
\centering
\includegraphics[width=2.6in]{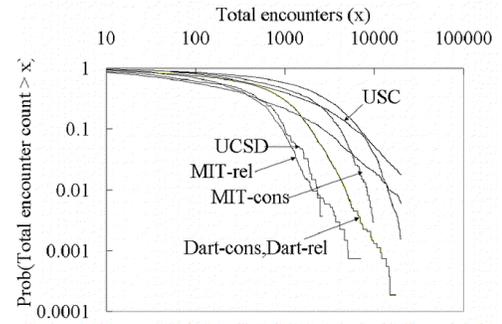}
\caption{CCDF of total encounter count}
\label{TE}
\end{figure}

We also show the CCDF of the total encounter events a MN has throughout the trace period in Fig. \ref{TE}. Again we observe significant difference in total encounter counts, another evidence of heterogeneous behavior among MNs. The actual number of total encounters depends on the size of population in the traces. Large traces (i.e., USC and Dartmouth traces) tend to have more encounters than small traces (i.e., UCSD trace). However, regardless of the trace population, the curves for total encounter count seem to follow BiPareto distribution \cite{bi-perato}. We try to fit BiPareto distribution curves to the empirical distribution curves, and use Kolmogorov-Smirnov test to examine the quality of fit. The resulting D-statistics for all traces are between $0.068$ and $0.024$, which indicates we have a reasonably good fit between the BiPareto distribution curves and the empirical distribution curves. For details about the Kolmogorov-Smirnov test and the parameters of fitted curve, please refer to appendix A.

\begin{figure}
\centering
\includegraphics[width=2.6in]{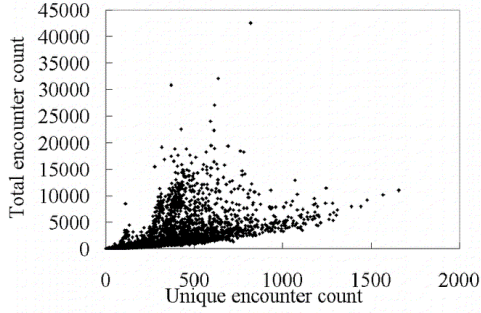}
\caption{Total encounter is not directly proportional to unique encounter (USC). Other traces show similar trend.}
\label{UEvsTE}
\end{figure}

A closer investigation into unique encounter count and total encounter count of the same MN reveals that high unique encounter count does not always imply high total encounter count, as shown in Fig. \ref{UEvsTE} using USC trace as an example. The correlation coefficient between unique encounter count and total encounter count is only $0.585$ in this example. This implies some node pairs have many repetitive encounters, suggesting closer relationship between such node pairs than others. In the next section, we introduce the notion of {\it friendship} between MNs to further understand such phenomenon.

{\it \textbf{In all the traces, the MNs encounter a relatively small fraction of the user population; below 40\% in most cases and never reaching above 60\% in any case. Except for UCSD trace, on average a MN only encounters $1.88\%$-$5.94\%$ of the whole population. The number of total encounters for the users follows a BiPareto distribution, the parameters of which depends on the campus.}}

\subsection{Friendship between Nodes} \label{friend}

In our daily lives, we are bound to meet with colleagues and friends much more often than others. In this section we try to investigate using the wireless LAN traces whether such uneven distribution of closeness among MN pairs exists, and try to measure it using the concept of {\it friendship dimensions}. The likelihood for encounters to occur for two given MNs captures the {\it friendship} between them. This "friendship" in WLAN trace may or may not reflect social friendship, which is impossible to validate from anonymized traces. We propose to identify friendship between MN pairs based on three different dimensions: Encounter duration, encounter count, and location diversity of encounter, with the following definitions:

\begin{itemize}
\item \textbf{Friendship based on encounter time}: We define friendship index based on duration of encounter as $Frd_{t}(A,B) = E_{t}(A,B)/OT(A)$, which is the ratio of sum of encounter durations between node $A$ and $B$, $E_{t}(A,B)$, to total online time of node $A$, $OT(A)$. This is an index for how good a friend node $B$ is to node $A$ based on duration of encounters. Note that in general $Frd_{t}(A,B) \neq Frd_{t}(B,A)$ and $0.0 \leq Frd_{t}(A,B) \leq 1.0$ for any node pair $A$ and $B$.

\item \textbf{Friendship based on encounter count}: The friendship index based on encounter count is defined as $Frd_{c}(A,B) = E_{c}(A,B)/S(A)$, which is the ratio between association sessions of node $A$ that contains encounter events with node $B$, $E_{c}(A,B)$, to total association session count of node $A$, $S(A)$.

\item \textbf{Friendship based on encounter location diversity}: The friendship index based on location diversity of encounter is defined as $Frd_{l}(A,B) = E_{l}(A,B)/L(A)$, which is the ratio between number of locations at which node $A$ has encounters with $B$, $E_{l}(A,B)$, to total locations node $A$ visits, $L(A)$.

\end{itemize}

The above three dimensions can be used to understand friendship between MNs from different perspective. Different relationships between MNs may lead to various friendship index value in these dimensions. For example, two users may have high friendship index based on encounter time and encounter count, but not from the location diversity perspective, and so on.

\begin{figure}
\centering
\includegraphics[width=2.6in]{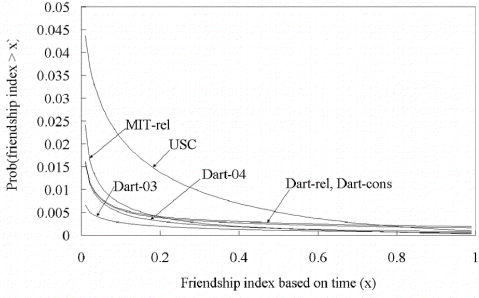}
\caption{CCDF of friendship index based on time}
\label{findex-time}
\end{figure}

We first observe how friendship indexes distribute among all ordered node pairs in the campuses studied. As shown in Fig. \ref{findex-time}, the CCDF curves of  friendship indexes based on encounter time follow exponential distributions for all campuses. Again we use Kolmogorov-Smirnov test to examine the quality of fit. The resulting D-statistics for all traces are between $0.0356$ and $0.0052$, which indicates we have a reasonably good fit between the exponential distribution curves and the empirical distribution curves. Please see appendix A for a brief introduction to Kolmogorov-Smirnov test and the detailed parameters of fitted exponential distribution curves. In spite of the fact that the traces are collected from different campuses with different methods, the shape of distribution curves of friendship index remain unchanged. We observe higher friendship index for the USC trace because the associations are measured at coarser granularity at switch port level, hence it is more likely for MNs to encounter one another.

To understand the effect of different trace collection methods on friendship index, we compare the distribution curve from Dart-04, Dart-rel and Dart-cons. The reconstructed traces tend to omit association sessions that are shorter than the sampling period, hence under-estimate the total online time for nodes, {\it OT(A)}, leading to slightly larger friendship indexes.

Exponential distribution of friendship index is an indication that majority of nodes do not have tight relationship with one another. In all the traces, only less than $5\%$ of ordered node pairs $(A,B)$ have friendship index $Frd_{t}(A,B)$ larger than $0.01$. This reveals the fact that in addition to limited fraction of nodes with encounter events as shown in the previous section, even for node pairs that do encounter with each other, most of them do not show tight relationship. Among all node pairs with non-zero friendship index, only $4.47\%$ of them has friendship index larger than $0.7$, and another $11.85\%$ of them with friendship index between $0.4$ to $0.7$. Friendship indexes based on encounter frequency or location diversity of encounter also show similar exponential distributions.

\begin{figure}
\centering
\includegraphics[width=2.6in]{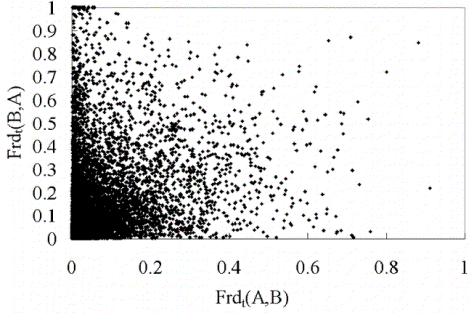}
\caption{Friendship indexes are asymmetric, high friendship index value for node pair (A,B) does not imply the same for (B,A). Shown for MIT trace. Other traces exhibit similar trend.}
\label{asymmetry-time}
\end{figure}

We next look into the issue of whether friendship index for an ordered node pair $Frd_{t}(A,B)$ and its reversed tuple $Frd_{t}(B,A)$ are symmetric. We plot the friendship index based on time for all node pairs with non-zero encounter duration in Fig. \ref{asymmetry-time} using MIT trace as an example. The scatter diagram shows that friendship indexes are highly asymmetric for each node pair. From Table \ref{friend-corr} we can see that for all three dimensions to define friendship indexes, the correlation coefficient between ordered node pair $(A,B)$ and $(B,A)$ are mostly low in all traces, implying high asymmetry in friendship indexes.

\begin{table}
\caption{Correlation coefficient for friendship indexes for all traces}
\label{friend-corr}
\begin{center}
\begin{tabular}{|c||c|c|c|}
\hline
 \mr{Trace name} & \multicolumn{3}{c}{Friendship index based on} \\
\cline{2-4}
 & encounter time & encounter count & location diversity \\
\hline
 MIT-rel & 0.415 & 0.327 & 0.186 \\
\hline
 UCSD & -0.024 & -0.004 & -0.003 \\
\hline
 USC & 0.158 & 0.205 & 0.130 \\
\hline
 Dart-03 & 0.351 & 0.278 & 0.043 \\
\hline
 Dart-04 & 0.629 & 0.201 & 0.068 \\
\hline
\end{tabular}

\end{center}
\end{table}

{\it \textbf{Friendship between MNs is highly asymmetric. The distribution for the friendship index is exponential for all the traces, regardless of the friendship definition (based on time, encouner, or location). Among all node pairs there are less than $5\%$  with friendship index larger than $0.01$, and less than $1\%$ with friendship index larger than $0.4$.}}

\subsection{Encounter-relationship graph} \label{ERG}

In section \ref{encounter}, we see that MNs have small percentage of unique encounters among the whole population. Given this fact, we raise a question regarding the possibility of establishing campus-wide relationships among majority of MNs via encounters alone. That is, do encounters link MNs on campus into one single community, or just small pieces of cliques?

To investigate this question, we define a static {\it encounter-relationship graph} (ER graph) as follows: Each MN is represented by a node in the {\it ER graph}, and an edge is added between two nodes if the two corresponding MNs have encountered at least once during the studied trace period. The concept of {\it ER graph} is introduced to capture potential for establishing relationships based on direct encounters.

We use three important metrics to describe the characteristics of encounter-relationship graphs, defined as follows:

\begin{itemize}
\item
\textbf{Clustering coefficient} (CC) is used to describe the tendency of nodes to from cliques in the graph. It is formally defined as:

$CC = \frac{\sum_{node i=1}^{M} CC(i)}{M}$

where $CC(i) = \frac{\sum_{A \in N(i)} \sum_{B \in N(i)} I(A\in N(B))}{Nbr(i) \cdot (Nbr(i)-1)}$

$I(\cdot)$ is the indicator function, $Nbr(i)$ is the number of neighbors node $i$ has, $N(i)$ is the set of neighbors of node $i$, and $M$ is the total number of nodes in the graph.

Intuitively, clustering coefficient is the average ratio of neighbors of a node that are also neighbors of one another.

\item
\textbf{Disconnected ratio} (DR) is used to describe the connectivity of ER graph. It is defined as:

$DR = \frac{\sum_{node A=1}^{M} \sum_{node B \notin C(A)} 1}{M(M-1)}$

where $C(A)$ is the set of nodes that are in the same connected sub-graph with node $A$.

\item
\textbf{Average path length} (PL) is used to describe the degree of separation of nodes in the {\it ER graph}. It is formally defined as:

$PL = (1-DR) \cdot PL_{con}+DR \cdot PL_{disc}$

Where $PL_{con}$ is the average path length among the connected part of the {\it ER graph}, defined as:

$ PL_{con} = \frac{\sum_{node A=1}^{M} \sum_{node B \in C(A)} PL(A,B)}{\sum_{node A=1}^{M} \sum_{node B \in C(A)} 1}$

$PL(A,B)$ is the hop count of the shortest path between node pair $(A,B)$. $PL_{disc}$ is the penalty on average path length for {\it disconnected} node pairs in {\it ER graph}. In the following we use the average path length of regular graphs (defined later) with the same node number and average node degree for $PL_{disc}$.

\end{itemize}

\begin{figure}
\centering

\includegraphics[width=2.6in]{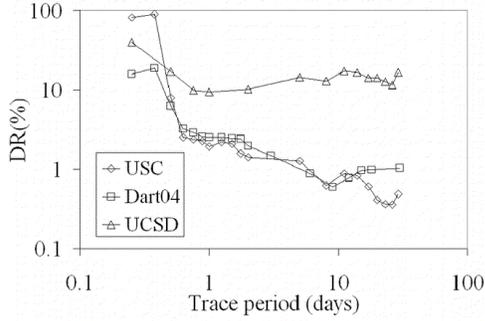} 

(a)Disconnected ratio

\includegraphics[width=2.6in]{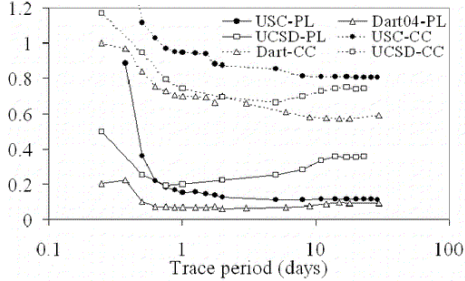}

(b)Normalized clustering coefficient and average path length



\includegraphics[width=2.6in]{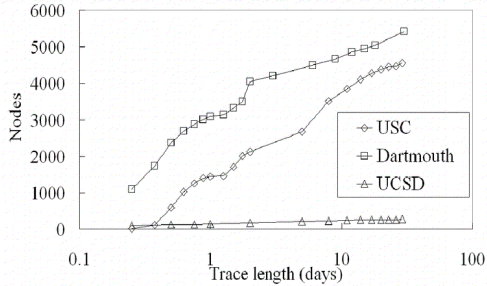} 

(c) Number of node in graph

\caption{Change in the {\it ER graph} metrics with respect to trace period}
\label{SmallWorld-evolve}
\end{figure}

We study how the above metrics evolve for the {\it ER graphs} derived from various studied period of WLAN traces. Taking USC trace, Dartmouth trace (Dart-04), and UCSD trace as examples, we show the evolution of the three metrics with respect to various studied trace periods in Fig. \ref{SmallWorld-evolve} (a)-(c). The graphs for other traces are not shown here due to limited space, but they also show very similar trends. The additional graphs are available in appendix B. To highlight a unique property of these {\it ER graphs}, we also calculate {\it CC} and {\it PL} for regular graphs and random graphs with the same corresponding total node number $M$ and average node degree $d$. In regular graphs, nodes are first arranged on a circle and each node is connected to $d$ closest neighbors on the circle. In random graphs, $d$ randomly chosen nodes are assigned as neighbors for each node. Typically, regular graphs have high {\it CC} and {\it PL} while random graphs have low {\it CC} and {\it PL}.


From Fig. \ref{SmallWorld-evolve} (a) we note that the {\it ER graphs} for generic wireless users (USC and Dart-04) have low {\it DR}, which implies that nodal encounters are sufficient to provide opportunities to connect almost all nodes in a single community, even though some of them are online for only short durations. This is an encouraging result that points out the feasibility of building a large, campus-wide relationship network relying only on direct encounters. For UCSD trace {\it DR} is higher. We suspect this is due to the total user number is comparitively small (only $275$ PDA users). We further point out that although {\it DR} starts out very high with very short trace period, as MNs have not moved around to create encounters yet, it decreases rather quickly as trace period increases. Within one day, {\it DR}'s reduces to within less than $3\%$ difference of its corresponding final values.

Another interesting finding is revealed by taking a further look at the other two metrics, clustering coefficient ({\it CC}) and average path length ({\it PL}). In Fig. \ref{SmallWorld-evolve} (b), we show the normalized {\it CC}'s and {\it PL}'s for various trace periods. These normalized metrics are obtained by the following equations:

$ CC_{norm} = \frac{CC-CC_{rand}}{CC_{reg}-CC_{rand}} 0.99 + 0.01 $

$ PL_{norm} = \frac{PL-PL_{rand}}{PL_{reg}-PL_{rand}} 0.99 + 0.01 $

where $CC_{norm}$ and $PL_{norm}$ represent normalized $CC$ and $PL$, respectively. The subscripts $reg$ and $rand$ imply the corresponding metric is obtained from the reguler graph and random graph, respectively, with the same total node number and average node degree. These normalized metrics represent, on the scale from $0.01$ (correspond to the random graph) to $1$ (correspond to regular graph), where do the metrics obtained from the {\it ER graph}s fall.

We observe that {\it ER graphs} display high {\it CC}'s which are close to those of corresponding regular graphs (i.e., Normalized {\it CC}'s being close to 1), and low normalized {\it PL}'s which are close to those of corresponding random graphs. This highlights that a special pattern of encounters exists in all network traces: Nodes having the same home AP are highly likely to encounter with all others and introduce highly connected clusters among these nodes, leading to high {\it CC}. Some of the nodes in one cluster also have random encounters with nodes in other clusters, and those links serve as "shortcuts" in the {\it ER graphs} that reduce {\it PL}. In previous literature, graphs with high {\it CC} close to regular graphs and low {\it PL} close to random graphs are referred as SmallWorld graphs \cite{Watts-SmallWorld}, \cite{Helmy-SmallWorld}. By looking at various traces, we indicate that the relationship formed by encounters among nodes using wireless network is also an instantiation of SmallWorlds. We also observe that both {\it PL} and {\it CC} converges to its final values rather quickly in about one day for USC and Dart04 traces. Although number of nodes in {\it ER graph} keep increasing as studied trace period increases, as shown in Fig. \ref{SmallWorld-evolve} (c), it does not change these metrics a lot.

{\it \textbf{Encounters link most of the MNs together in a connected graph, albeit each MN encounters only with small portion of the whole population. The encounter graph is a SmallWorld graph, and even for short time period its clustering coefficent, average path length, and connectivity are all close to those for longer traces.}}

\subsection{Encounter-relationship graph with friends} \label{ERG-friend}


In the previous section the {\it ER graph} is constructed by including all encounters to construct links between nodes. Typically, a MN may maintain relationship selectively only with those MNs that are considered "trust-worthy". For example, a MN may choose to trust those MNs with which it has high friendship indexes. The criteria of choosing the nodes to keep a relationship may influence the structure of {\it ER graphs}. This issue is the main focus of investigation in this section.

We defined the metrics for friendship in wireless networks in section \ref{friend}. Now we try to include friends with various degree of closeness in the {\it ER graph}, and see how it influences the structure of the graph. We use friendship index based on time as an example to show how choosing encounters with different degree of closeness can change the structure of {\it ER graph} significantly.

We sort the list of nodes that a node $A$ has encountered according to friendship index, $Frd_{t}(A,B)$, where $B$ is a node that encountered $A$ at least once. After sorting, each node decides to pick a certain percentage of nodes from the list with which to establish relationships. We choose nodes from top, middle, or bottom of the list and with various percentages, and obtain the corresponding metrics for the new {\it ER graphs} that include only the links to the chosen nodes.

In this case, one minor modification needs to be made to the metrics introduced earlier. Since friendship indexes are asymmetric as shown in section \ref{friend}, it is possible that node $A$ has chosen to include node $B$ in the {\it ER graph}, but not vice versa. Hence the modified {\it ER graph} becomes a directed graph instead of an undirected one. The definition of clustering coefficient is modified as follows:

$ CC = \frac{\sum_{node i=1}^{M} CC(i)}{M}$

where $CC(i) = \frac{\sum_{A \in F(i)} \sum_{B \in F(i)} I(A \in F(B))}{Frd(i) \cdot (Frd(i)-1)}$

$I(\cdot)$ is the indicator function, $Frd(i)$ is the number of friends node $i$ chooses to include in the graph, $F(i)$ is the set of chosen friends of node $i$, and $M$ is the total number of nodes in the graph. Note that $A \in F(B)$ does not imply $B \in F(A)$.

When calculating average path length and disconnected ratio, the paths must follow the direction of edges on the graph.

\begin{figure}
\centering
\includegraphics[width=2.6in]{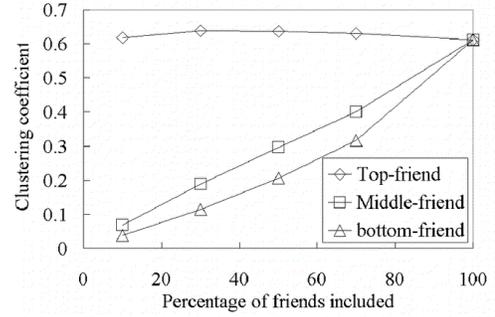}

(a) Clustering coefficient

\includegraphics[width=2.6in]{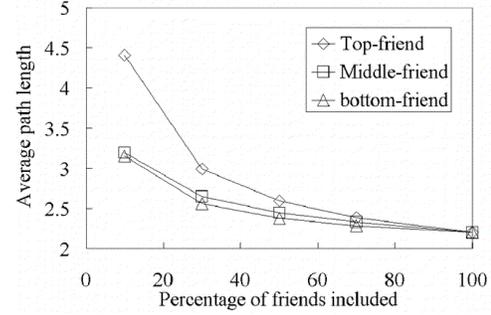}

(b) Average path length

\includegraphics[width=2.6in]{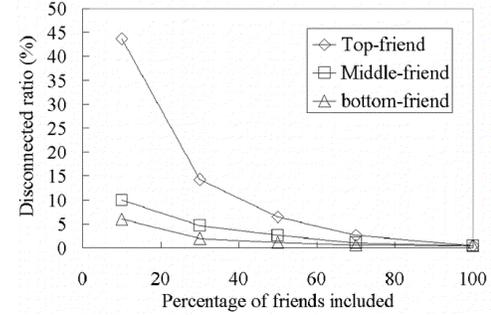}

(c) Disconnected ratio

\caption{Metrics of encounter-relationship graph by taking various percentage of friends}
\label{percent-friend}
\end{figure}

Following the modified definitions, we obtain the metrics when including given percentages of all encountered nodes from the top, middle, or bottom of the sorted encounter node list according to friendship index based on time. The figures are shown in Fig. \ref{percent-friend}. We use USC trace with 30-day trace duration as an example, and similar results are also observed in other traces.

The figures show a clear trend that if neighbors ranked high in friendship index are included, the resultant graph shows stronger clustering, and the average path length is much higher. The result stems from the fact that top friends of a given node are also likely to be top friend between one another, forming small cliques in the graph. Clustering coefficient remains high due to these cliques. Disconnected ratio and average path lengths are high due to the lack of links between different cliques. On the other hand, when low-ranked friends are included in the graph, the links included are distributed in a more random fashion, reflected by the low clustering coefficient and low average path length. Similar results are also observed in social science study of friendship between pupils \cite{social-study}. As larger portion of friends are included in the graph, all three metrics move closer to the value in section \ref{ERG} when all encounters are included.

We further perform the same experiment using the other two friendship index. the corresponding results are shown in appendix C for briefness.

{\it \textbf{Top-ranked friends tend to form cliques and low-ranked friends are the key to provide random links and reduce the degree of separation in encounter graph.}}

\subsection{Information diffusion using encounters}
\label{information_diffusion}

In addition to establishing relationship between nodes, encounters can also be utilized to diffuse information throughout the network. In this model, information is spreaded with nodal mobility and encounters, where nodes exchange information when they encounter each other directly. The speed and reachability of information diffusion among the nodes are determined by the actual patterns and sequences of encounters. In this section we seek to answer the question of whether the {\it current} encounter patterns between MNs in wireless networks are rich enough to be utilized for information diffusion. If the answer is yes, what is the delay incurred is such information diffusion scheme, and how robust is it?

As a first step to understand the problem, we use the simplest diffusion mechanism. We assume sufficient bandwidth and reliable communication between MNs, and sufficient storage space on all MNs. When a source node has information to send, it simply transmits it to all nodes it encounters if they have not received the information yet. All intermediate nodes cooperate in information diffusion, keeping a copy of received information and forwarding it the same way as the source node does. This simple approach is known as epidemic routing in the literature \cite{epidemic}. Under perfect environment with sufficient resources, it achieves lowest delay and highest delivery rate. Note that the delivered information may not take the shortest path in terms of transmission hop count, as all nodes simply take the earliest available chance to propagate the information. In this work we have chosen epidemic routing to show the potential of encounter-based information diffusion under {\it realistic} encounter patterns. We further investigate the robustness of the information diffusion scheme.

\begin{figure}
\centering
\includegraphics[width=2.6in]{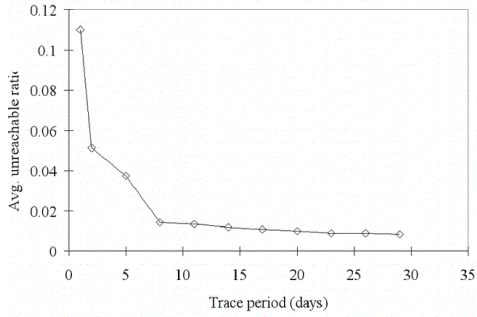}
\caption{Receive ratio of broadcast messages using epidemic routing}
\label{receive-ratio}
\end{figure}

In the following simulations, we use a traffic pattern in which the source node has some information it wishes to send to all other nodes. The source starts to "diffuse" the information when it is first online. As time evolves, nodes encounter with each other and increasing portion of the whole population receive the information. We study the percentage of nodes that received the information with various trace periods and show the results in Fig. \ref{receive-ratio}, using USC trace as an example. Each point in the figures of this section is an average value for using $30\%$ of the nodes that appear earliest in the corresponding trace period as sources.


From Fig. \ref{receive-ratio} we observe that even within a short trace period (i.e., one day) the information can be diffused to around $89\%$ of the whole population. As the trace period increases, reachability also improves. Given that most nodes are online for part of the trace period (Fig. \ref{CCDF-ontime}), visit only small portion of the whole campus (Fig. \ref{CCDF-coverage}), and encounter a small portion of the whole population (Fig. \ref{UE}), this result is perhaps beyond our original expectation. It gives a positive confirmation that it is potentially possible to deliver information relying only on encounters, in a campus environment with high success rate, under {\it current} user behavioral pattern. As the population of wireless computing devices and their average online time both increase in the future, we can expect to have even higher delivery rate. This subject bears futher research.

\begin{figure}
\centering
\includegraphics[width=2.6in]{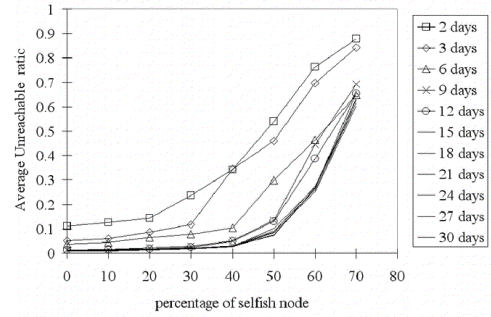}
\caption{Information delivery ratio with various selfish node percentage and trace period}
\label{selfish-DR}
\end{figure}

In some cases, a portion of nodes may not be cooperative to propagate the information, especially for a diverse user population as in university campuses. To understand how uncooprative users potentially influence the feasibility of information diffusion, we carry out the following experiment: We make a portion of users {\it selfish} such that it never forwards information for other sources, and we study the performance degradation under this setup. For each of the trace periods used, we increasingly make a certain percentage of nodes selfish, starting from those with highest unique encounter counts. By making nodes with high unique encounters selfish first, we eliminate more transmission opportunities than pick selfish nodes randomly, hence we expect to observe greater impact on performance. The relationship between percentage of selfish node and the information delivery ratio is shown in Fig. \ref{selfish-DR}. The result is very surprising: For all trace period tested, the unreachable ratio does not increase significantly before at least $20\%$ of nodes are selfish. The performance is even more robust if we take longer period of trace. This implies that even a significant portion of users are not willing to propagate information for others, the underlying nodal encounter pattern is rich enough for the information to find an alternative way through. Hence delivery rate is quite robust for up to an intermediate percentage of selfish nodes. It is also interesting to observe that for trace period longer than 15 days, the reachability curves almost overlap. This indicates that for trace periods longer than 15 days, few new encounters are introduced between node pairs. Most of encounters are simply repetitive ones between same pairs of nodes, hence longer trace period do not help to further improve robustness by introducing new potential paths.

\begin{figure}
\centering
\includegraphics[width=2.6in]{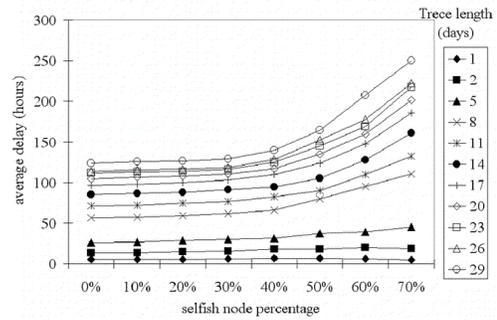}
\caption{Average message delay with various selfish node percentage and trace period}
\label{selfish-delay}
\end{figure}

We further show how average delay of information diffusion changes with increasing selfish node percentage in Fig. \ref{selfish-delay}. In the figure, average delay increases for longer trace duration because information that is not deliverable in shorter trace periods become deliverable. More interestingly, for all tested trace durations, average delay does not increase significantly before more than $40\%$ of nodes are selfish. This implies average delay is also robust against selfish user behavior up to an intermediate percentage.

{\it \textbf{In the current traces, encounters patterns are rich enough to support information diffusion. Specifically, information can be delivered to more than $94\%$ of users within two days. The reachability and average delay do not decrease significantly until at least $20\%-40\%$ of nodes are selfish.}}

\section{Discussions and future work} \label{disc}

In this section we discuss about the insights we gained from this work and point out potential future research directions.

\begin{itemize}

\item Modeling of users has always been an important research problem in all type of networks. In section \ref{user-metric} we propose metrics that fall in four categories to describe user behavior in WLAN traces. Among these, we emphasize that user on-off behavior, small coverage, and repetitive patterns are important features but they are largely overlooked by earlier work on mobility modeling \cite{mobility-model}, \cite{stationarity-model} and wireless network simulation. Previous works on wireless network user modeling \cite{WLAN-model}, while attempted to preserve user association durations, also did not attempt to preserve those important features. In the future we would like to work on a model that uses the dimensions proposed in section \ref{user-metric} to describe users in wireless LANs.

\item As a comparison between trace-based models and synthetic mobility models, we argue that low encounter percentage as show in section \ref{encounter} is not observed in any of the simulation scenarios used for performance evaluation in the literature. In typical synthetic mobility scenarios, all nodes follow the same model to make movement decision, albeit with randomness, and eventually encounter with all of other nodes \cite{MAID-tech}. The encounter pattern from real wireless network trace reflects that university campus is a heterogeneous environment rather than a homogeneous one constructed by synthetic mobility models. To better understand how protocols perform in such heterogeneous environment, using synthetic models would not be sufficient.

\item Although it is not possible to establish the exact reason behind the closeness of some MN pairs, this information may be utilized in several applications, such as better algorithms for cluster-forming in ad hoc networks, or finding a node to temporarily store a packet with higher probability to deliver it later to the final recipient. Protocols that are aware of social relationship among MNs may be an interesting direction in the future.

\item Generally, in social-relationship aware mechanisms, one tends to trust top-ranked friends more than the others. However, as we see in section \ref{ERG-friend}, using top-ranked friends only results in an {\it ER graph} with high clustering coefficient and average path length, and may lead to a disconnected relationship network. In order to remain connected to a larger community, one should also use some randomly-chosen users (or middle friends) to reduce the degree of separation in underlying {\it ER} graph.

\item The information diffusion model we proposed (in Section~\ref{information_diffusion} can be related to several new research directions. Recently, the delay tolerant network (DTN) paradigm has been proposed to deliver messages in highly dynamic networks where network partitions are frequent~\cite{DTN}. In order to reduce overhead, an important issue in DTN protocols is to select the next hop intelligently. In the future, we intend to work on a message-forwarding mechanism based on our analysis of encounter patterns. 

Our robustness and delay analysis of information diffusion over encounter graphs show two interesting points: (1) For message delivery, the delivery ratio and delay are not affected significantly, even if we can not choose the shortest paths due to non-cooperative users. (2) On the other hand, it would be difficult to prevent diffusion of harmful or malicious messages, such as computer worms or viruses from propagating through encounters. Both observations are due to the richness in underlying encounter pattern providing multiple chances for message delivery.

\end{itemize}

\section{Conclusion} \label{conclusion}

In this paper we study the wireless network traces from four different university campuses collected by various methods, with focus on different user populations. To the best of our knowledge, this is the most comprehensive study to date on wireless LAN traces in the literature.

We first propose metrics to describe individual user behavior, pointing out important common features in all studied traces that are not emphasized by previous works. Wireless network users on university campuses are characterized by large percentage of offline time, limited visited APs on the campus, and repetitive association patterns. We believe that these metrics capture important characteristics about users in wireless networks and should be included in user modeling for more accurate performance evaluation in the future. We also find the detailed distributions are different from studied traces, due to the difference in underlying user population and trace collection methods.

We further study the relationship between MNs in these traces. We find that encounters and friendship are asymmetrically distributed among all MNs, indicating that the user population is a heterogeneous one. Using encounter-relationship graph, we establish that it is possible to create a campus-wide community based solely on nodal encounters. The relationship graph of such a community can be described using SmallWorld graphs. Finally, we use epidemic routing with trace-based encounter pattern to show that information diffusion is feasible in current wireless networks, and its performance is robust to an intermediate proportion of selfish users.

The metrics and experiments we use in the paper provide a further step towards understanding of user behavior in wireless network, which provide good foundation for future research.

\section*{Acknowledgments}

This work was funded in part through the NSF CAREER Award 0134650.

We would like to thank Brian Yamaguchi and Carl Hayter at USC ISD division for helping us collected the WLAN traces on the USC campus for the past 2 years. We would also like to extend our thanks to David Kotz, Marvin McNett, and Magdalena Balazinska for helping us obtain the Dartmouth, UCSD, and MIT traces, respectively. We are also thankful to USC HPCC for providing high quality computing facility. In addition, we would like to thank Fan Bai and Shao-cheng Wang for providing useful feedback on this work.

\section*{Appendix A. BiPareto distribution and Kolmogorov-Smirnov test}
In this section we first briefly introduce Kolmogorov-Smirnov test and biPareto distribution, and then list the detail numerical results of using biPareto distribution curves to fit total encounter distributions obtained in section \ref{encounter}.

BiPareto distribution is first used in \cite{bi-perato} to fit the number of connections per user TCP session and mean connection inter-arrival time in a TCP session. Later, BiPareto distribution is again used in \cite{NCCH-trace} to fit the distribution of association session length in wireless LAN. In this work, we use it to fit the distribution of encounters a MN has in WLANs. The CCDF of BiPareto distribution is as follows:

$Prob(X>x) = (\frac{x}{k})^{-\alpha}(\frac{x+c}{k+c})^{\alpha-\beta},\: x>k$

$Prob(X>x) = 1,\: x \leq k$

The left part of CCDF curve of BiPareto distribution on log-log scale is a straight line with slope $-\alpha$. As the $x$ variable comes close to the turning point, $c$, the slope of the CCDF curve gradually changes from $-\alpha$ to $-\beta$. In our study of total encounter distributions, we choose $k=1$ for all curves.

Kolmogorov-Smirnov test is used to determine whether the hypothesized distribution (in our case, the BiPareto ditribution) adquately fit the empirical distribution. K-S test is not sensitive to the binning of data set, unlike Chi-square test. Therefore we choose K-S test in our study.

\begin{figure}
\centering
\includegraphics[width=2.6in]{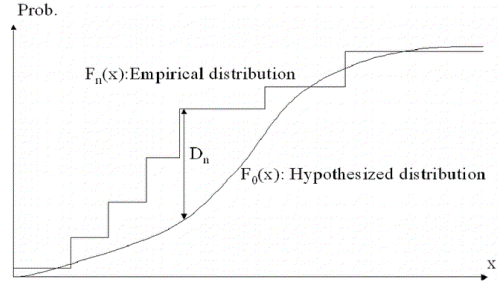}
\caption{Illustration of D-statistics and K-S test}
\label{K-S}
\end{figure}

Referring to Fig. \ref{K-S}, in K-S test the distances between the hypothesized distribution and the empirical distribution are measured at all $x$ values, and the maximum of the measured distances is called D-statistics. More formally, D-statistics is defined as:

$D_{n} = sup_{x}[\mid F_{n}(x) - F_{0}(x) \mid ]$

where $F_{n}(x)$ and $F_{0}(x)$ are the empirical and hypothesized distributions, respectively. Intuitively, D-statistics measure the maximum difference between the two distribution curve. A smaller D-statistic indicates a better fit of the hypothesized distribution to the empirical distribution.

We use minimum squared error method to find the best fit of BiPareto distribution curves to the empirical total encounter distributions for various traces. The parameters are listed in Table \ref{K-S table}. From the table we observe that the D-statistics are no larger than $0.05$ except for UCSD trace, for which D-statistic is still reasonably low at $0.07$, indicating a reasonable fit of the BiPareto distribution.

\begin{table}
\caption{BiPareto distribution fitting to total encounter curves and D-statistics for K-S test}
\label{K-S table}
\begin{center}
\begin{tabular}{|c||c|c|c|c|}
\hline
 \mr{Trace name} & \multicolumn{3}{c}{BiPareto parameters} & \mr{D-statistics} \\
\cline{2-4}
 & $\alpha$ & $\beta$& $c$ & \\
\hline
 MIT-rel & 0.027 & 9.8 & 4000 & 0.036 \\
\hline
 MIT-cons & 0.029 & 3.0 & 4500 & 0.040\\
\hline
 UCSD & 0.062 & 16.3 & 9900 & 0.068\\
\hline
 USC &  0.019 & 0.83 & 550 & 0.049 \\
\hline
 Dart-03 & 0.0723 & 0.81 & 290 & 0.049 \\
\hline
 Dart-04 & 0.0285 & 4.43 & 11850 & 0.025 \\
\hline
 Dart-rel & 0.037 & 7.46 & 7200 & 0.031\\
\hline
 Dart-cons & 0.037 & 30.4 & 30900 & 0.024 \\
\hline
\end{tabular}
\end{center}
\end{table}

In section \ref{friend} we use exponential distribution to fit the empirical distributions of friendship index. The CDF of exponential distribution is given by:

$F(x) = 1-e^{-\lambda x}, \; \; x \geq 0$

We list the $\lambda$ parameters we obtained using minimum squared error method to fit exponential distributions to the empirical distribution of friendship indexes based on encounter time in table \ref{friend K-S}. The corresponding D-statistics are also listed.

\begin{table}
\caption{Exponential distribution fitting to friendship index based on encounter time curves and D-statistics for K-S test}
\label{friend K-S}
\begin{center}
\begin{tabular}{|c||c|c|}
\hline
 Trace name & $\lambda$ & D-statistics \\
\hline
 MIT-rel & 369.19 & 0.0167 \\
\hline
 USC & 305.3 & 0.0356 \\
\hline
 Dart-03 & 500.4 & 0.0052 \\
\hline
 Dart-04 & 411.81 & 0.0116 \\
\hline
 Dart-rel & 409.91 & 0.0120\\
\hline
 Dart-cons & 412.35 & 0.0119\\
\hline
\end{tabular}
\end{center}
\end{table}

\begin{figure}
\centering

\includegraphics[width=2.6in]{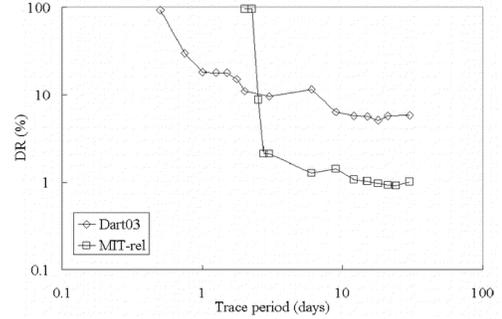} 

(a)Disconnected ratio

\includegraphics[width=2.6in]{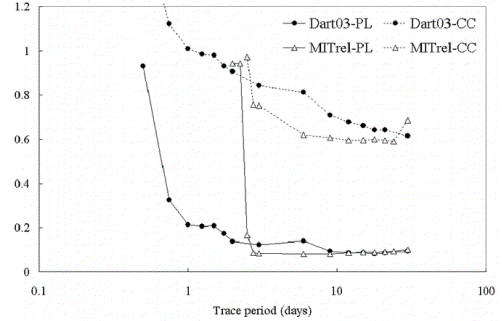}

(b)Normalized clustering coefficient and average path length



\includegraphics[width=2.6in]{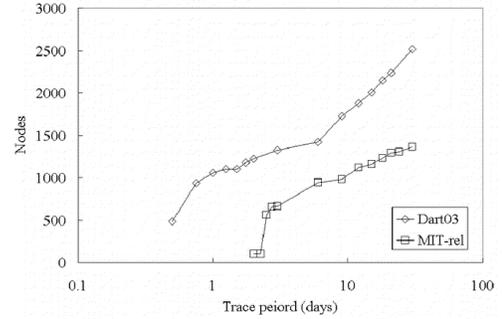} 

(c) Number of node in graph

\caption{Change in the {\it ER graph} metrics with respect to trace period}
\label{SmallWorld-evolve-app}
\end{figure}

\section*{Appendix B. Additional graphs for encounter-relationship graphs metrics}

In addition to the figures shown in section \ref{ERG}, we also obtain the same metrics for MIT-rel and Dart-03 traces. The corresponding figures are shown in Fig. \ref{SmallWorld-evolve-app}, displaying similar trends as discussed in section \ref{ERG}. One interesting observation here is that for MIT trace, disconnected ratio is very high until day 3 in the trace. A further investigation reveals that MIT trace collection was started on a Saturday, and for a pure working environment Saturdays and Sundays are the least active days. The disconnected ratio is almost $100\%$ until day 3 because the MNs that are on during the weekend are mostly stationary ones. We observe a jump of number of node in the trace, a sudden decrease in {\it DR}, and an abrupt change in both {\it CC} and {\it PL} on day 3.

\section*{Appendix C. Encounter-relationship graphs metrics based on various friendship indexes}

In this section we show the clustering coefficient, average path length, and disconnected ratio if we use different percentage of friends, using friendship indexes based on encounter count and encounter location diversity, in the {\it ER graph} generated from USC trace. The definition of the friendship indexes and the metrics are given in section \ref{friend} and \ref{ERG-friend}, respectively.

\begin{figure}
\centering
\includegraphics[width=2.6in]{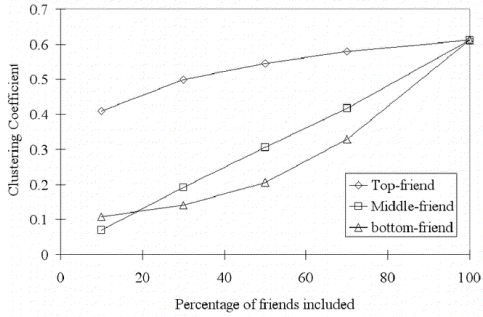}

(a) Clustering coefficient

\includegraphics[width=2.6in]{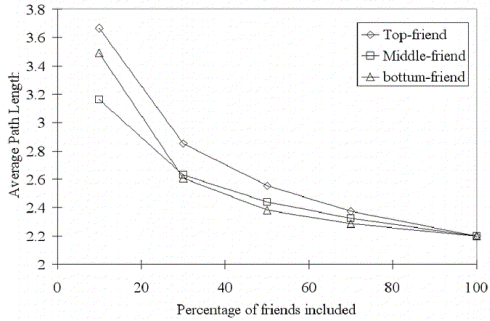}

(b) Average path length

\includegraphics[width=2.6in]{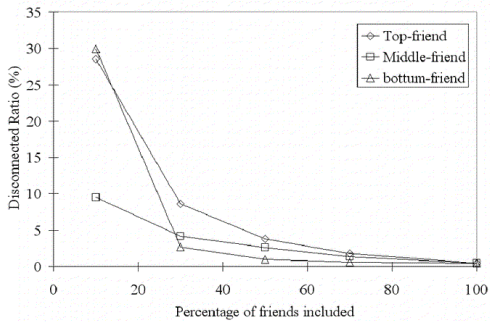}

(c) Disconnected ratio

\caption{Metrics of ER graph by taking various percentage of friends based on encounter count}
\label{percent-friend-count}
\end{figure}

Fig. \ref{percent-friend-count} shows how the metrics change by including different percentage of friends based on encounter count. The trend is similar to what we observe in Fig. \ref{percent-friend}, where we use friends based on encounter time. Using top friends tend to lead to small cliques and hence high {\it CC}, {\it PL}, and {\it DR} in the generated {\it ER graph}. On the other hand, lower ranked friends represent the random links in {\it ER graph}, leading to low {\it CC}, {\it PL}, and {\it DR}.

\begin{figure}
\centering
\includegraphics[width=2.6in]{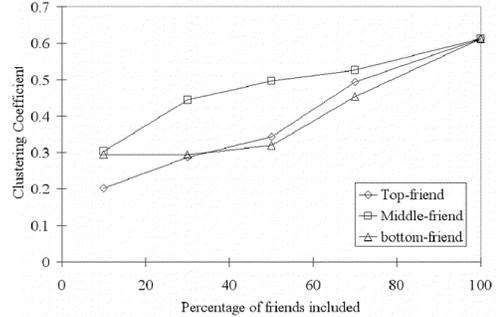}

(a) Clustering coefficient

\includegraphics[width=2.6in]{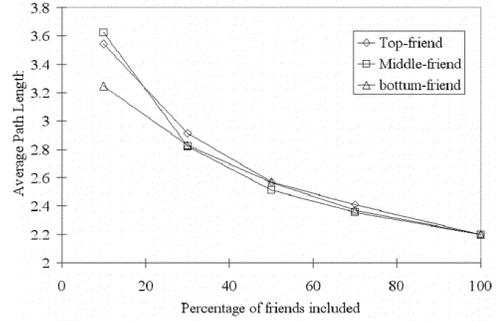}

(b) Average path length

\includegraphics[width=2.6in]{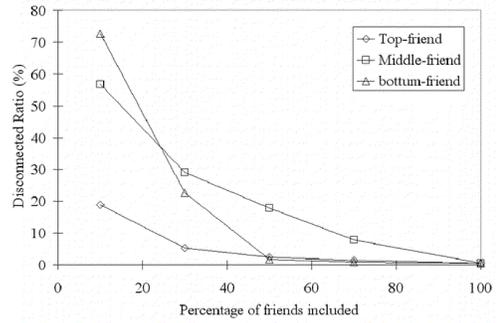}

(c) Disconnected ratio

\caption{Metrics of ER graph by taking various percentage of friends based on encounter location diversity}
\label{percent-friend-loc}
\end{figure}

However, this trend is not clear if we use friends based on encounter location diversity. As shown in Fig. \ref{percent-friend-loc}, the curves of using top, middle, or bottom ranked friends cross each other, and there is no clear trend as opposed to the graphs in which the other two friendship indexes are used. The reason is that since most MNs do not visit many APs, the friendship index based on encounter location diversity is a less effective way to distinguish the actual degree of friendship. On the extreme, for a MN that only visits one AP, all the encounters it has must occur at this AP, and the friendship index based on encounter location diversity is $1.0$ for all MNs it encounters. Therefore, picking top, middle, or bottom ranked friends would degenerate to randomly picking friends, resulting in less obvious trend in the {\it ER graph} metrics when we pick friend based on encounter location diversity.

\end{document}